\documentclass[epsfig, 12pt]{article}
\usepackage{makeidx}
\usepackage{epsfig}
\usepackage{mathrsfs}
\usepackage{latexsym}
\usepackage{amsfonts}
\usepackage{amssymb}
\usepackage{amsmath}
\usepackage{amsthm}
\input epsf.sty
\begin{document}

\title{%
\rightline{\mbox {\normalsize
{}}\bigskip}\textbf{ Generalized Nariai Solutions for Yang-type Monopoles     }}
\author{ Pablo  Diaz\thanks{pdiaz@unizar.es}, 
Antonio  Segui\thanks{segui@unizar.es}\\
{\small  Departamento de Fisica Teorica, Universidad de Zaragoza, 50009-Zaragoza, Spain.}
} \maketitle
\begin{abstract}
A detailed study of the geometries that emerge by a gravitating generalized Yang monopole in even dimensions is carried out. In particular, those which present black hole and cosmological horizons. This two-horizon system is thermally unstable. The process of thermalization will drive both horizons to coalesce. This limit is what is profusely studied in this paper. It is shown that eventhough coordinate distance shrinks to zero, physical distance does not. So, there is some remaining space which geometry has been computed and identified as a generalized Nariai solution. The thermal properties of this new spacetime are then calculated. Topics, as the elliptical relation between radii of spheres in the geometry or a discussion about whether a mass-type term should be present in the line element or not, are also included.

\end{abstract} 

\newpage

\tableofcontents

\section{Introduction} 

Monopoles have been subject of deep study and controversy all over the last century. This is so because, although no experimental evidence of their existence has been found, many theoretical issues make them almost unavoidable. They already appeared as solutions of Maxwell equations as long as the null $B$-divergence condition was relaxed, that is, $\nabla\cdot B\ne 0$. It was Dirac~\cite{D} in the early thirties who first proposed the theoretical possibility of creating an experiment to actually produce a ``fake'' monopole, in a way that its fakeness, say, the Dirac string, was undetectable. As a consequence, the product of the electric and the magnetic charges was quantized. Many years later, in 1959, the quantization requirement was confirmed by the celebrated Aharonov-Bohm experiment~\cite{AB}.

Since 1954, owing to the papers by Yang and Mills~\cite{YM} and by Utiyama~\cite{U}, gauge theories of a group of symmetry larger than $U(1)$, in particular non abelian symmetry groups $SU(2)$ and $SU(3)$ (which eventually would conform the Standard Model of particle physics) where gradually developped. In 1969, Lubkin~\cite{L} realized that monopoles can be classified by the homotopy group of the gauge symmetry group of the theory, so that the magnetic charge is replaced by the topological charge of the field configuration. In the case of the Dirac monopole, the homotopy group $\pi_{1}$ of $U(1)$ is exactly $Z$. However, it was not till 1975 that Yang~\cite{Y} generalized the abelian monopole to the case of an $SU(2)$-invariant gauge theory in six dimensions, see also~\cite{Nuyts}. 
Modern approaches use the formalism of fiber bundles for a suitable description of monopoles. It generalizes the traditional classification in terms of the homotopic group of the gauge theory. In this way, magnetic monopoles are identified with the different instanton configurations which come up basically as non trivial maps of the gauge group, usually $SU(N)$, onto $S^{d}$, where $d$ is the spatial dimension. That is, magnetic monopoles are all those non trivial principal bundles with group structure $SU(N)$ that can be realized on the hypersurface $S^d$. The classification coincides, as said before, with the different classes of homotopy groups.
The genalization of Yang monopoles to an arbitrary even dimension was carried out in~\cite{HP}. 
Using slightly different methods similar analysis have recently been done~\cite{GT}.
The reader can find good reviews on the subject in~\cite{C},~\cite{WY} and the references therein.

As every existing object in nature, monopoles couple to gravity via their energy-momentum tensor. The resulting geometry is obtained by solving the Yang-Mills-Einstein equations, which get greatly simplified by imposing spherical symmetry (as expected from a magnetic monopole field configuration). This geometry is fully specified by choosing a point in the space of parameters $\{\mu, m, \Lambda, k \}$, the meaning of which will be explained in detail later on. For a given range of parameters, it is easy to prove that the geometry presents both a cosmological and an event horizon. A full analogy with the Schwarzschild-de Sitter solution reveals that, in these cases, the geometry is dynamically driven through the parameter space into a thermally stable point where both horizons coalesce~\cite{GP}, the final line element being the analogue of Nariai's spacetime in four dimensions.

This paper is organized as follows: the next section sets a general framework and fixes the notation used later. The main body of the article concerns the analysis of the coalescence solutions. This is achieved in two subsections corresponding to the massless and massive cases respectively. An explicit computation of the resulting geometry is carried out in each case. A final section includes some conclusions and comments. Two appendixes have been added to the article. They are topics which lie somehow out of the main line of the paper, either for being technical aspects of a computation (Appendix A) or for presenting a new idea the exposition of which would need a new section, as in Appendix B. the absence of B, in turn, would not have prevented the reader from a full understanding of the paper.

\section{The gravitational coupling. Some geometrical features}
The gravitational effects of these monopoles have been recently studied ~\cite{GT}. It was done, as usual, by minimally coupling the Yang-Mills energy-momentum tensor to gravity. Variations of the Einstein-Hilbert action
\begin{equation}
S = \int dx^d \sqrt{-\det g}\,  \left[ {1\over 16\pi G} \left(R -2\Lambda\right)
- {1\over2\gamma^2}\, Tr |F|^2  \right]\
\end{equation} 
with respect to the metric tensor leads to
\begin{equation}\label{eq:EA}
G_{mn}=8\pi G T_{mn}-g_{mn}\Lambda,
\end{equation}
where 
\begin{equation}\label{eq:EMtensor}
T_{mn}=\gamma^{-2}\bigg[\mathrm{tr} (F_{m}^{\phantom{m}p} F_{np})-\frac{1}{4}g_{mn}\mathrm{tr} (F_{pq}F^{pq})\bigg]
\end{equation}
is the energy momentum tensor of the YM strength field. The traces are taken in the colour index and $\gamma$ is the YM coupling constant. Finding general solutions for (\ref{eq:EA}) is a highly complicated problem. 
However, imposing spherical symmetry simplifies the task enormously. According to this, the ansatz will be a spatially spherically symmetric $(2k+2)$-dimensional metric whose line element reads
\begin{equation}\label{eq:metric}
ds^{2}=-\Delta dt^{2}+\Delta^{-1} dr^{2}+r^{2}d\Omega^{2}_{2k}.
\end{equation}
The last equation is consistent with (\ref{eq:EA}) and (\ref{eq:EMtensor}) when ~\cite{GT}
\begin{equation}\label{eq:delta}
\Delta(r)=1-\frac{2Gm}{r^{2k-1}}-\frac{\mu^{2}}{r^{2}}-\frac{r^{2}}{R^{2}},
\end{equation}
where $R=\sqrt{\frac{k(2k+1)}{\Lambda}}$ is the de Sitter radius, $\mu^2$ is proportional to $\frac{1}{2k-3}$ and measures the magnetic charge of the monopole, $m$ comes up as a constant of integration with dimensions of mass and $G$ is the Newton constant in $2k+2$ spacetime dimension.
At first sight, (\ref{eq:metric}) with (\ref{eq:delta}) look like a Schwarzschild-de Sitter geometry in $2k+1$ spatial dimensions with an extra term, the one involving $\mu$, which seems to be independent of the dimension of spacetime. It seems reasonable to think of this term as a contribution of the magnetic monopole. This simple image, even if not exact\footnote{The resulting geometry is, of course, not just the sum of terms of different geometries, but it casually coincides. Differences are bound to exist on the limit of vanishing of a given contribution. For instance, let us suppose that, given a set of parameters, say \{$m,\mu, \Lambda$, k \}, we can switch off $\mu$ (by neglecting it with respect to the others). The resulting geometry is topologically different to the one obtained by not assuming any monopole at all at the beginning, that is, the limit does not coincide. However, in the cases studied here, this is no more than an enough-to-be-aware-of subtlety.}, is helpful and, unless we face the vanishing limits, it may be kept in mind in the following. 

The next step (and the next temptation) is to analyze how the causal structure of this spacetime depends on given values of the parameters. The main body of this work concerns a deep analysis of the solution in the case when parameters $\mu, \Lambda$, $m$ and $k$ allow the existence of two horizons. Then, inspired by the  Schwarzschild-de Sitter unstable solution, it is claimed that the system gets dynamically driven to a value of the parameters where both horizons coalesce. Eventhough coordinate distance shrinks to zero, physical distance does not. A generalized Nariai geometry ``between'' the horizons is then explicitly obtained.
The Nariai line element~\cite{N} is a nonsingular solution of the Einstein's vacuum equations with a positive cosmological constant, $R_{\mu \nu}=\Lambda g_{\mu \nu}$. It was first found by Kasner~\cite{K} and its electrically charged generalization dates of 1959~\cite{B}. However, the important fact that it emerges as an extremal limit of Schwarzschild-de Sitter black holes was not noticed until 1983~\cite{GP}.

Nariai spacetime in four dimensions is the direct product $dS_{2} \times \mathcal{S}^{2}$, $dS_{2}$ being no more than the hyperbolic version of $S^{2}$ as we change $t \to i\tau$. In $2k+2$ dimensions, the solution gets generalized to $dS_{2} \times \mathcal{S}^{2k}$. Again, it is the direct product of two constant curvature spaces and admits a $3+k(2k+1)$ group of isometries $SO(2,1) \times SO(2k+1)$. The space is homogeneous since the group acts transitively and is {\it locally} static, given that a global $dS$-type spacetime cannot be described by merely one static coordinate chart. In four dimensions, radii of curvature of the two product spaces are equal if the black hole is neutral, and different in the charged case. If the black hole is electrically charged, the respective radii $a$ and $b$ are different and related by the equation 
\begin{equation}\label{eq:ab}
a^{-2}+b^{-2}=2\Lambda
\end{equation}
as shown in~\cite{B}. This relation will be generalized in the magnetic case, the object of our study. A short but instructive recent work on the four dimensional geometry can be found in~\cite{O}.  

\section{The horizon coalecence geometry}
Studying the horizons of a geometry like (\ref{eq:metric}) is equivalent to searching the divergencies of $g_{rr}$ for finite values of the coordinates. This leads us to analyze the zeroes of function $\Delta(r)$, where the horizons will be located. For a certain range of values of $\{\mu, \Lambda, k, m \}$ there will be two horizons. Finding this region in the parameter space will be the first task.
After that, attention will be focused on the coalescence point of the horizons\footnote{Coalescence as seen in Schwarzschild coordinates.}. The analysis consists of two steps, first, the parameterization of the coordinate separation of the horizons ($\epsilon $) and the calculation of the physical distance between them when coalescence takes place ($\epsilon \to 0$). Then, following the strategy in~\cite{GP}, the computation of the line element of the remaining geometry. This program is carried out on two cases: $m=0$ and $m\ne0$, which are treated in the next subsections, respectively. The massless case must be seen as a toy model of the massive one. This distinction is made not merely for simplicity  but also because, as will be explained, the mass parameter comes out naturally for dynamical requirements. 
\subsection{Case $m=0$}
In the massless case, $\Delta(r)$ gets reduced to 
\begin{equation}\label{eq:mld}
\Delta(r)=1-\frac{\mu^{2}}{r^{2}}-\frac{r^{2}}{R^{2}}.
\end{equation}
Solving $\Delta=0$ is equivalent to finding the zeroes of a biquadratic equation as long as $r=0$ is not considered. We perform the change $z\equiv r^{2}$ and solve a second order ordinary equation. The horizons are found to be at
\begin{eqnarray}
z_{+}&=&\frac{R^{2}}{2}\bigg(1-\sqrt{1-\frac{4\mu^{2}}{R^{2}}}\bigg)\\
z_{++}&=&\frac{R^{2}}{2}\bigg(1+\sqrt{1-\frac{4\mu^{2}}{R^{2}}}\bigg). 
\end{eqnarray} 
$R>2\mu$ guarantees the existence of two positive solutions and, therefore, four solutions for the quartic equation. Two of them, $r_{+}=+\sqrt{z_{+}}$ and $r_{++}=+\sqrt{z_{++}}$, correspond to the radial coordinate of the inner (black hole) and outer (cosmological) horizon respectively. If $R=2\mu$, both solutions coincide, which means that the horizons coalesce. As said before, this does not mean that the geometry vanishes as a naive observation (given a wrong choice of coordinates) would make one think. Physical distance between the horizons, on the contrary, remains finite at the limit. In order to prove this, let us compute it. For fixed time and angular coordinates, the physical distance is
\begin{eqnarray}
D(\mu,R)&=&\int_{r_{+}}^{r_{++}}\frac{rR}{[-r^{4}+R^{2}r^{2}-\mu^{2}R^{2}]^{1/2}}dr={}\nonumber\\ & & {} =\int_{z_{+}}^{z_{++}}\frac{R}{2\Big[\Big(\frac{R^{4}}{4}-\mu^{2}R^{2} \Big)-\Big(z-\frac{R^{2}}{2}\Big)^{2}\Big]^{1/2}}dz
\end{eqnarray} 
The requirement $R>2\mu$ implies $\frac{R^{4}}{4}-\mu^{2}R^{2}>0$ so the above integral is exactly solved as an $\cos^{-1}$-type. The result is
\begin{equation}
D(R)=\frac{1}{2}\pi R.
\end{equation}
Surprisingly, the physical distance does {\it not} depend on $\mu$. It means that, given a cosmological constant $\Lambda$, one could ``switch on'' the monopole and go on till the horizons coalesce but the distance would remain unchanged. However, because of quantization requirements, monopole charge $\mu$ cannot be tuned, but needs to have, instead, a fixed value upto a sign. On the other hand, the cosmological constant, $\Lambda$, should be chosen when writing the lagrangian. It means that changing its value does not drive us from one model to another but implies an essential change in the theory~\cite{GH}. Therefore, we are not free to adjust any parameter arbitrarily as done with the mass of the black hole in the Schwarzschild-de Sitter case. Then, eventhough physical reasons would lead the horizons to coalesce, the absence of any free parameter in our model makes it impossible. In the next section, $m$ will come to our help as a free parameter for the model.

Despite the last remark, one could wonder about the kind of geometry that remains when the horizons coalesce. This task, even if seems just a curious exercise now, will be useful for the next section. Applying a technique similar to the one Gingspar and Perry~\cite{GP} used to study the geometry of Nariai's solution, we proceed by, first, parametrizing the separation of horizons as
\begin{equation}\label{eq:separation}
R=2\mu(1+\epsilon^{2}),
\end{equation}
in a way that coalescence corresponds to taking $\epsilon=0$. Then, we define a ``wise'' change of coordinates
\begin{eqnarray}\label{eq:coordinates}
\chi &=& \cos^{-1}\Big[\frac{-2}{R^{2}A}(r^{2}-r_{0}^{2})\Big] \nonumber\\
\tau &=& \epsilon \frac{2it}{R},
\end{eqnarray}    
where $A=\sqrt{1-\frac{4\mu^{2}}{R^{2}}}$ and $r_{0}^{2}=\frac{R^{2}}{2}$, and the angular coordinates remain unchanged. The new coordinates (\ref{eq:coordinates}) might seem randomly chosen at first sight. However, there are some reasons that justify such a functional dependence. For instance, $\chi$ is nothing but the physical distance between $r_{+}$ and $r$. The timelike coordinate $t$ is multiplied by $i$ in order to work in the Euclidean region\footnote{$\tau$ will be periodic at both horizons, although different in each case. Equality will hold at the coalescence point, when thermal stability is reached.} and by $\epsilon$ because $\Delta/\epsilon^{2}$ is expected to have a finite limit when $\epsilon\to 0$. Now, we apply (\ref{eq:separation}) and (\ref{eq:coordinates}) and expand $\Delta(r(\chi))d\tau^{2}$, $\Delta^{-1}(r(\chi))d\chi^{2}$ and $r^{2}(\chi)$ up to first order in $\epsilon$. The line element (\ref{eq:metric}) 
reads    
\begin{eqnarray}
ds^{2} &=& \mu^{2}d\chi^{2}+\mu^{2}\sin^{2}(\chi)\big[1+\epsilon\sqrt{2}\cos(\chi)\big]d\tau^{2}+{}\nonumber\\ & &{}+2\mu^{2}\big[1-\sqrt{2}\cos(\chi)\epsilon\big]d\Omega_{2k}^{2}.
\end{eqnarray}
We take limit $\epsilon\to 0$ to obtain
\begin{equation}\label{eq:nar2k}
ds^{2}=\mu^{2}\big[d\chi^{2}+\sin^{2}(\chi)d\tau^{2}\big]+2\mu^{2}d\Omega_{2k}^{2}.
\end{equation}
As seen in (\ref{eq:nar2k}), the $2k$-sphere decouples from the rest. The resulting geometry is $S^{2}\times S^{2k}$ for $k\ge2$. Notice the parallelism between this geometry and Nariai's solution, which is $S^{2}\times S^{2}$. The ``classical'' relation between radii (\ref{eq:ab}) gets also generalized to
\begin{equation}\label{eq:berg}
a^{-2}+b^{-2}=C_{0}\Lambda,
\end{equation}  
where $C_{0}=\frac{6}{k(2k+1)}$.
The geometry (\ref{eq:nar2k}) can be viewed as a ``degenerate'' black hole, in which the two horizons have the same (maximum) size and are in thermal equilibrium. This could be interpreted by an observer as a bath of radiation coming from both horizons at a precise temperature~\cite{BH}. The temperature can be calculated by means of surface gravity $\kappa$, as computed in the new coordinates (\ref{eq:coordinates}) 
\begin{equation}\label{eq:temp1}
T=\frac{1}{2\pi \sqrt{2} \mu}=\frac{1}{\pi}\bigg[ \frac{k(2k+1)}{2\Lambda}\bigg]^{1/2}.
\end{equation}
The entropy can also be computed as a quarter of the sum of the two horizons~\cite{HR}, so
\begin{equation}\label{eq:entropy1}
S=\frac{1}{4}\mathcal{A}_T=\frac{1}{2}\mathcal{A}_H=\frac{1}{2}\omega_{2k}\bigg[\frac{k(2k+1)}{2\Lambda}\bigg]^{2},
\end{equation}
where $\omega_{2k}$ is the area of the $2k$-dimension unit sphere.
\subsection{Case $m\ne 0$}\label{sec:massive}
In the massive case we recover the full expression (\ref{eq:delta}) for $\Delta$. Since the singular point $r=0$ is not to be considered, we better analyze the function $r^{2k-1}\Delta(r)$
\begin{equation}\label{eq:md}
\widetilde{\Delta} \equiv r^{2k-1}\Delta=-\frac{r^{2k+1}}{R^{2}}+r^{2k-1}-\mu^{2}r^{2k-3}-2Gm.
\end{equation}
It is known that a polynomial equation with powers equal to or higher than five is not generally solvable in a symbolical way. This happens for $k \ge 2$. So, the purpose of doing a study for the massive case analogous to that achieved in the first section is ruined. Nevertheless, some information can be extracted from (\ref{eq:md}). We should first remember the sign of the parameters: $R^2>0$ (de Sitter), $\mu^{2} >0$ for $k\ge 2$, and $m$ will be free in principle. Derivating (\ref{eq:md}) and equating to zero leads to a biquadratic equation of the form
\begin{equation}\label{eq:biquadratic}
-\frac{1}{R^{2}}(2k+1)r^{4}+(2k-1)r^{2}-(2k-3)\mu^{2}=0,
\end{equation}
which, as long as 
\begin{equation}\label{eq:bound}
\Lambda\mu^{2}\le\frac{k}{4}\frac{(2k-1)^2}{2k-3},
\end{equation}
has two positive (and two negative) roots, $r_{min}$ and $r_{max}\equiv r_{c}$. In terms of the cosmological constant 
\begin{equation}\label{eq:rc}
r_{c}^{2}\equiv \frac{k(2k-1)}{2\Lambda}\Bigg[1+\sqrt{1-\frac{4(2k-3)\Lambda \mu^{2}}{k(2k-1)^{2}}}\Bigg],
\end{equation}
$r_{min}$ is obtained from (\ref{eq:rc}) by swapping the sign of the square root. A quick look at (\ref{eq:md}) shows that the smallest root is a minimum and the largest is a maximum of function $\widetilde{\Delta}$. Now, let us plug $r_{c}$ into (\ref{eq:md}):
\begin{enumerate}
 
\item If $m>0$, then (see fig.1)
\begin{itemize}
\item[a)] $\widetilde{\Delta} (r_{c})\ge 0$ implies that there are two event horizons, the black hole and the cosmological horizon. The inequality gets saturated at the coalescence point. 
\item[b)] $\widetilde{\Delta} (r_{c})<0$ means that no horizon is found.
\end{itemize}
\item If $m<0$, then (see fig.2)
\begin{itemize}
\item[a)] $\widetilde{\Delta}(r_{min})<0$ together with $\widetilde{\Delta}(r_{c})<0$ implies that there is just one Cauchy horizon.    
\item[b)] $\widetilde{\Delta}(r_{min})<0$ together with $\widetilde{\Delta}(r_{c})>0$ assures the existence of a Cauchy horizon and both black hole and cosmological horizon.
\item[c)] $\widetilde{\Delta}(r_{min})>0$ leaves us with the cosmological horizon only.
\end{itemize} 
\end{enumerate}
The case we will study is $\widetilde{\Delta} (r_{min})<0$ and $\widetilde{\Delta} (r_{c})>0$ which, independently of the sign of $m$, assures\footnote {The value of $m$ can be negative. That is because $m$ should not be thought of as an entity with physical meaning but as a geometrical parameter. Short calculation in (\ref{eq:m}) shows that $m$ gets negative values for $\Lambda\mu^{2}\ge \frac{k}{4}(1+2k)$.} the existence of black hole and cosmological horizons. This corresponds to values of $m$ within range (see fig.3)
\begin{figure}[h]
\centerline{\epsfig{file=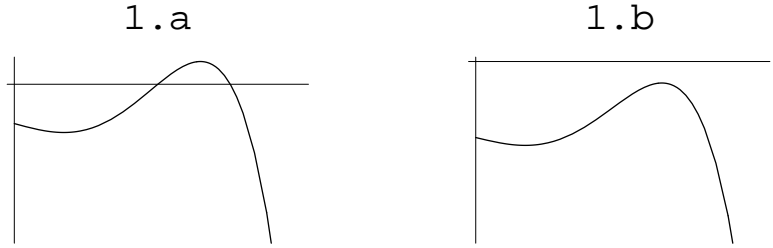, height=4cm}}\caption{{\small Case $m>0$. The curve represents function $\widetilde{\Delta}(r)$. Figure 1.a has two roots which correspond to the black hole ($r_{+}$) and cosmological horizon ($r_{++}$) respectively. Figure 1.b shows the absence of horizons.}}
\end{figure}

\begin{figure}[h]
\centerline{\epsfig{file=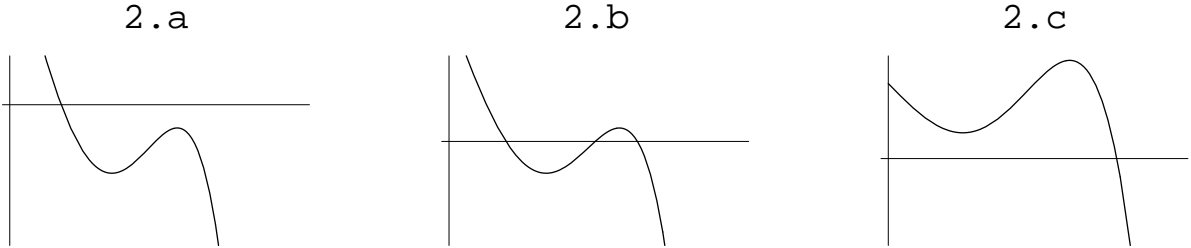, height=4cm}}\caption{{\small Case $m<0$. This time $\widetilde{\Delta}(r)$ permits the existence of one (Cauchy) horizon as in Figure 2.a, three horizons (Cauchy, black hole and cosmological) as in 2.b, or just the cosmological horizon as shown in 2.c.}} 
\end{figure}

\begin{figure}[h]
\centerline{\epsfig{file=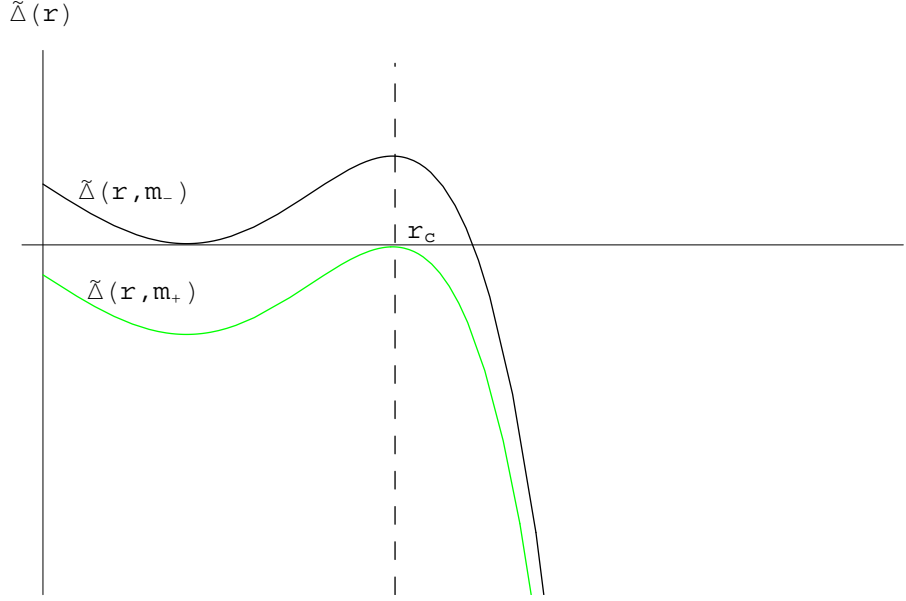, height=8cm}}\caption{{\small This figure shows the range of ``masses'' which are consistent with the existence of both black hole and cosmological horizons. The curve $\widetilde{\Delta}(r)$ ``moves down'' in the process of coalescence.}}
\end{figure}

\begin{equation}
m_{-}<m<m_{+},
\end{equation}
where
\begin{equation}\label{eq:mc}
Gm_{c}\equiv Gm_{+}=\frac{1}{1+2k}r_{c}^{2k-3}(r_{c}^{2}-2\mu^{2}).
\end{equation}
The value of $Gm_{-}$ is obtained by replacing $r_{c}\to r_{min}$. In terms of $\Lambda$ and $\mu$ we get
\begin{eqnarray}\label{eq:m}
G m_{\pm}&=&\frac{(2\Lambda)^{-k+1/2}}{1+2k}\Big[-k+2k^{2} \pm \sqrt{k^{2}(1-2k)^{2}-4\Lambda\mu^{2}(2k-3)k}\Big]^{k-3/2}{}\nonumber\\
& & {} \Big[-k+2k^{2}-4\Lambda\mu^{2} \pm \sqrt{k^{2}(1-2k)^{2}-4\Lambda\mu^{2}(2k-3)k}\Big].
\end{eqnarray}

The crucial point is that both horizons coalesce when $r_{c}$ is a root of (\ref{eq:md}) which happens at $m=m_{c}(k,\Lambda\mu^2,\Lambda)$. Two relations have been imposed so far: $\frac{d \widetilde{\Delta}(r;m)}{dr} \lvert_{r_{c}}=0$, that is, (\ref{eq:biquadratic}), which defines $r_{c}$, and $\Delta(r_{c};m_{c})=0$ which leads to $m_{c}$.
In order for $m$ to be real, the bound which must be impossed on $\Lambda\mu^{2}$ coincides with (\ref{eq:bound}) which, in turn, is nothing but the condition for the existence of two horizons. So, if a given a value for $\Lambda\mu^{2}$ is low enough to produce two horizons, there always exists a real value of $m$ which makes them coalesce. Again, as in the Schwarzschild-de Sitter example, the system is unstable and the equilibrium point is reached at $m=m_{c}$. Unlike the massless case, plugging $m$ gives us enough room for maneuvre to drive the system to equilibrium. \\
At this point, we would like to remark that the procedure of horizon coalescence, as studied in detail below, may be seen as a flow in a line which undergoes a Pitchfork bifurcation at the coalescence point. Parameter $m$, moved by thermal instability, drives the system to the critical situation. For concreteness see Appendix B.\\
Let us focus on the near coalescence point. This can be parameterized by
\begin{eqnarray}\label{eq:proxi}
r&=&r_c +\delta r=r_{c}(1+\epsilon\cos \chi)\\ \nonumber
m&=&m_{c}-\delta m=m_{c}(1+b\epsilon^{2}).
\end{eqnarray}
Parameterization of $r$ also involves a change of coordinates $r \to \chi$ and should be taken as imposed at the moment although it will be justified later. The horizons will be symmetrically located at: $r_{+}=r_{c}(1-\epsilon)$ and $r_{++}=r_{c}(1+\epsilon)$ which correspond to $\chi_{+}=\pi$ and $\chi_{++} =2\pi$, respectively\footnote{For a small enough $\epsilon$, it is expected that the parabolic approach holds and, then, both horizons are symmetrically located with respect to $r_{c}$.}. 
The value of $b$ as well as the absence of a linear term in $\epsilon$ of the parameterization of $m$ may be explained as follows. Near the coalescence point one should Taylor expand $\Delta$  around $r_{c}$ and have in mind that, for $\frac{\delta m}{m_{c}} \ll 1$, $\Delta$ is aproximately parabolic, so that second order expansion is enough. By definition $\Delta(r_{+})=\Delta(r_{++})=0$ and $\Delta$ reaches a maximum at $r_{c}$. So,
\begin{eqnarray}\label{eq:parabolic}
0&=&\Delta(r_{++})=\Delta(r_{c},m)+\Delta'(r_{c},m)(r_c \epsilon)+\frac{1}{2}\Delta''(r_{c},m)(r_c \epsilon)^{2}=\nonumber\\ &=&\frac{2G\delta m}{r_{c}^{2k-1}}+\frac{1}{2}\Delta''(r_{c};m_{c})r_{c}^{2}\epsilon^{2},
\end{eqnarray}
which means that 
\begin{equation}\label{eq:b}
b=\frac{\Delta''(r_{c};m_{c})r_{c}^{2k+1}}{4Gm_{c}}.
\end{equation} 

Calculating the physical distance near the coalescence point would, again, imply solving the integration
\begin{equation}\label{eq:dis}
D(\epsilon)=\int_{r_{+}}^{r_{++}}\frac{dr}{\Delta^{1/2} (r)},
\end{equation} 
where $r_{++}=r_{+}+2r_{c}\epsilon $. Although the exact result is not computed, an explicit proof of its finite nonzero value is given in Appendix A. The procedure of calculating the physical distance also brings us some light on which is the change of coordinates that should be made in order to understand the resulting geometry. It turns out to be 
\begin{eqnarray}\label{eq:newcoor}
\chi &=& \cos^{-1}\bigg[\frac{1}{\epsilon r_{c}}(r-r_{c})\bigg] \nonumber\\
\tau &=& \frac{\epsilon i}{B r_{c}} t,
\end{eqnarray}  
where 
\begin{equation}\label{eq:B}
B=\frac{k}{k-2k^{2}+2\Lambda r_{c}^{2}} 
\end{equation}
is a dimensionless factor.

The coalescence of horizons takes place at $\epsilon=0$. In order to study the geometry at the limit we proceed by calculating $-\Delta dt^{2}$, $\Delta^{-1} dr^{2}$ and $r^{2}$ in the new coordinates (\ref{eq:newcoor}) and expand in $\epsilon$ around $\epsilon=0$. The new line element gets determined by taking the zero order of the expansion. The relations for $r$ and $m$ in (\ref{eq:proxi}) are in accordance with (\ref{eq:newcoor}), where $b$ takes the value of (\ref{eq:b}), by virtue of the parabolic approach. 
From (\ref{eq:newcoor}), it is straightforward to see that $r^{2}$ takes a constant value $r_{c}^{2}$. Surprisingly, as in the massless and Schwarzschild-de Sitter cases, the geometry splits in two disconnected parts which lead to a product manifold $\mathcal{S}^{2}\times \mathcal{S}^{2k}$.
The line element reads
\begin{equation}\label{eq:geo}
ds^{2}=B r_{c}^{2}\Big[d\chi^{2}+\sin^{2} (\chi) d\tau^{2}\Big]+r_{c}^{2}d\Omega_{2k}^{2},
\end{equation}
where $\chi \in [\pi,2\pi]$ and $\tau$ is periodic\footnote{$\tau$ is periodic on both horizon surfaces all over the process in order to avoid the conical singularity at the horizons. At the coalescence point, however, both periods equal.}. 

As seen in (\ref{eq:geo}), $\mathcal{S}^{2}$ has radius $a^2=B r_{c}^{2}$, and $\mathcal{S}^{2k}$ has radius $b^2=r_{c}^{2}$. Now, the generalized Bertotti relation (\ref{eq:ab}) is 
\begin{equation}\label{eq:berge}
a^{-2}+b^{-2}=\frac{2(1-k)}{r_{c}^{2}}+\frac{2\Lambda}{k}=C\Lambda,
\end{equation}
where $C(k,\Lambda\mu^2)$ is obtained by inserting (\ref{eq:rc}) in (\ref{eq:berge}). 
Note that $C\big(k, \Lambda\mu^2=k(2k+1)/4\big)=C_{0}$, and then (\ref{eq:berge}) turns into (\ref{eq:berg}), that is, into the massless case. This is no surprising since $\Lambda\mu^2=k(2k+1)/4$ is the condition for coalescence in the massless case (equivalent to $R=2\mu$), and, at the same time, it makes $m_{c}=0$. So, the massive geometry is a consistent extension of the massless one. Now, fixing $\Lambda$ does not determine uniquely the geometry. Another dimensionless variable $\Lambda\mu^{2}$ is required.  

As in the last section, the geometry (\ref{eq:geo}) can be viewed as a ``degenerate'' black hole, in which the two horizons have the same (maximum) size and are in thermal equilibrium. In the present case  the temperature is given in terms of the surface gravity $\kappa$ by
\begin{equation}\label{eq:temp2}
T=\frac{\kappa}{2\pi}=\frac{1}{2\pi r_{c} \sqrt{B}}.
\end{equation}
In Planck units,the entropy associated with this solution may be calculated (given that it is not extreme\footnote{A charge black hole is said to be extreme when it has the minimum mass. Then, as it cannot release any energy without losing charge, it is supposed not to emit, and its associated Hawking temperature is 0. The black hole we are dealing with in this paper is extreme in the sense of carrying the ``maximum mass'' allowed by the cosmological constant $\Lambda$. Obviously, the temperature will not be zero.}) by means of the total area of the horizons as 

\begin{equation}\label{eq:entropy2}
S=\frac{1}{4}\mathcal{A}_T=\frac{1}{2}\mathcal{A}_H=\frac{1}{2}\omega_{2k} r_{c}^{4}.
\end{equation}

\section{Conclusions}
The spherically symmetric solution of gravity due to a magnetic monopole in arbitrary dimension has been studied, in particular, when the set of parameters $\{\Lambda, \mu, m, k\}$ allows the existence of two horizons. In these cases, thermal instabilities drive a process of horizon coalescence. Even though coordinate separation between the horizons shrinks to zero, it has been proven in both the massless and the massive case that the physical distance does not. The geometry of the remaining space between the horizons has been calculated in both cases. They turned out to be Nariai-type solutions, that is, the product of a 2-sphere and a $2k$-sphere for a $(2k+2)$-dimensional spacetime. In each solution, the radii of the spheres are not independent. They are related by an elliptical equation which should be understood as the generalization of the relation found by Bertotti. The unique generalized equation involving these radii for both the massless and the massive case has been given. After computing the line element in each case, the thermodynamical properties (Hawking temperature and entropy) due to the existence of horizons have been calculated.

The Yang monopole corresponds to the six dimensional case, where $k=2$. The geometry obtained after coalescence is $\mathcal{S}^{2}\times\mathcal{S}^{4}$ as can be explicitly read in (\ref{eq:geo}). This case is especially interesting since it may be described in String Theory (a realization of the Yang monopole in Heterotic String Theory has recently been done~\cite{BGT} as well as another complementary picture in Type-IIA String Theory~\cite{BDS1}). In the same context, it looks possible to find results (\ref{eq:entropy1}) and (\ref{eq:entropy2}) for the entropy by application of some attractor mechanism~\cite{FKS,FK}. We believe that this would be an interesting topic to be addressed in future research.     
  
\appendix
\section{Proof of the finite nonzero physical distance}
Computing the physical distance is equivalent to performing the integration
\begin{equation}\label{eq:dis}
D=\int_{r_{+}}^{r_{++}}\frac{dr}{\Delta^{1/2} (r)},
\end{equation} 
where, for small $\epsilon$, $r_{++}=r_{+}+2r_{c}\epsilon $. Divergencies might appear at the points where $\Delta \to 0$. The case we have been considering all along section (\ref{sec:massive}) concerns the existence of two horizons which coalesce, that is, two single roots $r_{+}$ and $r_{++}$ of $\Delta$ which join to form a double one. Function $\widetilde {\Delta}$ can always be expressed as $\widetilde {\Delta}=(r-r_{+})(r_{++}-r)g(r)$, where $g(r)$ is a polynomial function of powers of degree ${2k-1}$ and no zeroes within the range $[r_{+},r_{++}]$ are to be found by construction. Explicitly, equation (\ref{eq:dis}) is 
\begin{equation}
D(\epsilon)=\int_{r_{+}}^{r_{+}+2r_{c}\epsilon}\frac{dr}{(r-r_{+})^{1/2}(r_{+}+2r_{c}\epsilon-r)^{1/2}}\underbrace{\bigg(\frac{r^{k-1/2}}{g^{1/2}(r)}\bigg)}_{h(r)}.
\end{equation}
Now, $h(r)$ is a continuous divergenceless strictly positive function in the compact $[r_{+},r_{++}]$, which means that it will reach a positive maximum and minimum for certain $r's$. Let us call $h_{max}$ and $h_{min}$ the values of the function $h$ in these points\footnote{These, in principle, depend on $\epsilon$ but coincide when $\epsilon \to 0$: $h_{min}=h_{max} \equiv h(r_{c})$.}. Then
\begin{eqnarray}
h_{min}\int_{r_{+}}^{r_{+}+2r_{c}\epsilon}\frac{dr}{(r-r_{+})^{1/2}(r_{+}+2r_{c}\epsilon-r)^{1/2}} \le D(\epsilon) \le \nonumber\\ 
\le h_{max}\int_{r_{+}}^{r_{+}+2r_{c}\epsilon}\frac{dr}{(r-r_{+})^{1/2}(r_{+}+2r_{c}\epsilon-r)^{1/2}}.
\end{eqnarray}  
The integration can be performed:
\begin{equation}\label{eq:integration}
\int_{r_{+}}^{r_{+}+2r_{c}\epsilon}\frac{dr}{(r-r_{+})^{1/2}(r_{+}+2r_{c}\epsilon-r)^{1/2}}=\pi. 
\end{equation}
Now 
\begin{equation}
D(\epsilon \to 0)=\pi h(r_{c}), 
\end{equation}
where the value of $r_{c}$ is given in (\ref{eq:rc}).\\
Integrations of form (\ref{eq:integration}) are solved exactly by a $\cos^{-1}$ type function, and a nonzero finite result is obtained. It is remarkable that the same can be said for any $\Delta$ we would choose, as long as no more than two single roots were to join to form a double one. The key point is that (\ref{eq:integration}), which could be problematic, is independent of $\epsilon$ and therefore the distance is finite in the limit, when $\epsilon \to 0$. So, eventhough (\ref{eq:integration}) was neither exactly the physical distance in the massive case nor in Schwarzschild-de Sitter solution (however, it was in the massless case as we have already seen in the first section), it is closely related to it. This fact gives us a hint or, at least, justifies the change of coordinates we were performing once and again to study the geometry at the limit $\epsilon \to 0$. 
\section{Horizon coalescence as a flow on the line}
The main phenomenon that concerns this paper, as said before, can be described in terms of the dynamics of a vector field on the line. The coalescence point, in this picture, is no more than a supercritical Pitchfork bifurcation.
Let us remember some general features of the dynamics of a one-dimensional flow. The equation of a general vector field on the line can be expressed as:
\begin{equation}\label{eq:flow}
\dot{x}=f(x,\alpha)
\end{equation}
where $f$ is any real function with real support, the dot means differentiation with respect to $t$ and $\alpha$ is a parameter of the model. Fixed points of (\ref{eq:flow}) require $\dot{x}=0$, which must be obtained by finding the roots of $f$, that is
\begin{equation}\label{eq:fp}
f(x^{*},\alpha)=0.
\end{equation}
Equation (\ref{eq:fp}) is solved by an $n$-collection of fixed points $x_{i}^{*}$ for a given value of $\alpha$. Let us suppose that $f$ has three roots if $\alpha=\alpha_{0}$. Fixed points come closer as $\alpha $ moves and get ``condensed'' in a ``fat'' fixed point (bifurcation point) at $\alpha=\alpha_{c}$.  A paradigmatic example of a Pitchfork bifurcation is shown by function 
\begin{equation}\label{eq:pitchfork}
f(x)=x(\alpha-x^{2}).
\end{equation}
One question arises naturally now about the role the horizons play in this picture. Let us claim that horizons are fixed points and the role of $\alpha$ is played by $m$. We will justify this identification by constructing the vector flow.

Constructing a flow in a manifold (in our case it will be a line) is equivalent to giving a family of curves $\bar{r}(t)$ which covers the manifold or part of it. Each of the curves gets specified by the initial condition, say, $\bar{r}(t=0)$. Now, let us consider geodesic motions. Without loss of generality, the angular coordinates of our geometry will be frozen, $\theta$ and $\phi$ are constants, and only radial curves $r(t)$ are to be regarded. Static coordinate system will serve us to describe the movement for any $r\in (r_{+},r_{++})$. Let us invoke intuition at this point. If $r(0)$ is near the cosmological or the black hole horizon it is clear that a test particle will move out of the region by approaching each horizon respectively. Then, there is a point $r=r_{g}$ where the test particle will not ``feel'' any force and, consequently, it will not move\footnote{$r_g$ in our geometry, plays the role the asymptotic infinity does in Schwarzschild solution,that is, the point where the time-like Killing vector should be normalized in order to define the horizon temperature. Note that $r_{g}\equiv r_{c}$ at the coalescence point, that is, when $\epsilon=0$.}. This is the first (unstable) fixed point. 

Let us move the origin by defining $r'=r-r_{g}$, after this, primes will be dropped out to simplify notation. The flow at each point will be determined by the physical velocity $\dot{r}(r)$ (as measured by an observer placed at $r=0$) that a test particle would adquire at $r$ if it is dropped with $\dot{r}=0$ at around $r=0$ (as close as possible). It is not hard to see that the velocity of the test particle, as seen by the static geodesic observer, is bound to be zero at both horizons. So, horizons are fixed points. Now, our system can be treated as a vector flow $\dot{r}(r)$ which covers the region between the horizons. The vector flow has three fixed points: $\{r_{+}, r_{++}, 0\}$ where the first two are stable. As $m$ runs towards $m_{c}$, the system shrinks into a Pitchfork bifurcation. Near the bifurcation point the flow can be approximated by
\begin{equation}\label{eq:dotr}
\dot{r}=\beta r(r-r_{+})(r_{++}-r),
\end{equation} 
where $\beta$ is a positive constant which depends on $\mu$, $k$ and $\Lambda$. On the one hand, in the coordinate system $\{\chi, \tau\}$, and using (\ref{eq:newcoor}), we have
\begin{equation}
\frac{dr}{dt}=\frac{dr}{d\chi}\frac{d\chi}{d\tau}\frac{d\tau}{dt} \longrightarrow \dot{r}=\frac{i \epsilon^{2}}{B}\frac{d\chi}{d\tau}.
\end{equation}
On the other hand, equation (\ref{eq:dotr}), expressed in the new coordinate system, reads $\dot{r}=\epsilon^{3} \beta r_{c}^3 \cos \chi \sin^2 \chi$, and so
\begin{equation}
\frac{d\chi}{d\tau}=-i\epsilon \beta r_{c}^3 B \cos\chi \sin^2 \chi.
\end{equation} 
As expected, in the new coordinate system, every point converts into a fixed point as horizons coalesce ($\epsilon \to 0$). Since the flux lines were identified with geodesics of test particles, this can be understood as the abscence of forces at the end of the process. 
\section*{Acknowledgment}
We thank P. K. Townsend and Adil Belhaj for helpful discussions and Jean Nuyts for critical reading of the manuscript. This work has been supported by MCYT ( Spain) under grant FPA 2003-02948.

\end{document}